\documentclass[12pt]{article}
\usepackage[numbers]{natbib}
\usepackage{a4}
\usepackage{amsmath}
\usepackage{amssymb}
\usepackage{latexsym}
\usepackage{amssymb}
\usepackage{epsfig}
\usepackage{natbib}
\def \d{{\mathrm{d}}}

\def \R{{\mathbb{R}}}

\def \pd{\partial}

\def \Bs{\boldsymbol{s}}
\def \Bz{{\boldsymbol{z}}}

\def \BR{{\boldsymbol{R}}}
\def \BV{{\boldsymbol{V}}}

\def \BL{{\boldsymbol{L}}}
\def \BJ{{\boldsymbol{J}}}

\def \rr{{\boldsymbol{r}}}
\def \RR{{\boldsymbol{R}}}

\def \BB{{\boldsymbol{B}}}
\def \BE{{\boldsymbol{E}}}
\def \BH{{\boldsymbol{H}}}
\def \BD{{\boldsymbol{D}}}

\def \VV{{\cal{V}}} 
\def \tr{{t_{\text{r}}}}

\begin{document}
\title{{\bf 
The electromagnetic fields and the radiation of a spatio-temporally varying
electric current loop
}}
\author{
Markus Lazar~$^\text{a,b,}$\footnote{
{\it E-mail address:} lazar@fkp.tu-darmstadt.de (M.~Lazar);
\newline
Tel.:+49(0)6151/163686; Fax.: +49(0)6151/163681.
}
\\ \\
${}^\text{a}$ 
        Heisenberg Research Group,\\
        Department of Physics,\\
        Darmstadt University of Technology,\\
        Hochschulstr. 6,\\      
        D-64289 Darmstadt, Germany\\
${}^\text{b}$ 
Department of Physics,\\
Michigan Technological University,\\
Houghton, MI 49931, USA
}

\date{\today}    
\maketitle

\begin{abstract}
The electric and magnetic fields of a spatio-temporally varying
electric current loop are calculated using the Jefimenko equations. 
The radiation and the nonradiation parts of the electromagnetic fields
are derived in the framework of Maxwell's theory of electromagnetic fields. 
In this way, a new, exact, analytical solution of 
the Maxwell equation is found.
\\

\end{abstract}

\section{Introduction}
A current loop is an important item in the theory of electromagnetic 
induction and magnetostatics~\citep{PP,Jackson,Griffiths,HM}. 
Usually, such considerations are restricted to the 
static case or quasistatic case. No single work 
has been considered the general case of the non-uniform motion of an
arbitrary current loop or the electromagnetic radiation
fields of such a loop in the general framework 
of Maxwell's theory of electromagnetic fields, until now.

One important feature of the theory of electrodynamics
is the retardation which is a 
consequence of the finite speed of the propagating electromagnetic fields.
There is always a time delay 
since an effect observed by the receiver at the present position 
and present time was caused by the sender at some earlier time (retarded time) 
and at the retarded position.

The aim of this paper is the study of a spatio-temporally varying 
(or time-depending) closed current loop and the corresponding radiation fields.
In particular we investigate the so-called electrokinetic field.
The exact solutions of all the electromagnetic fields
induced by a non-stationary electric current loop are determined.
The Li\'enard-Wiechert fields produced by the electric current loop
are calculated.
The generalized Faraday law and the generalized 
Biot-Savart law for a spatio-temporally varying electric current loop
are found. 
The electromagnetic fields are decomposed into radiation and nonradiation
parts and finally, it is shown that the classical expression for
an electric wire is recovered as static limit from our general expressions.

\section{Basic framework}

The basic electromagnetic field laws are represented by the 
inhomogeneous and homogeneous Maxwell equations~\citep{Jackson,Griffiths}
\begin{align}
\label{ME-inh}
&\nabla\cdot \BD=\rho\,,\qquad \nabla\times\BH-\pd_t\BD=\BJ\,,\\
\label{ME-h}
&\nabla\cdot \BB=0\,,\qquad \nabla\times\BE+\pd_t\BB=0\,,
\end{align}
where $\BD$ is the electric displacement vector (electric excitation), 
$\BH$ is the magnetic excitation vector, 
$\BB$ is the magnetic field strength vector, $\BE$ is the electric field
strength vector, $\BJ$ is the electric current density vector, and 
$\rho$ is the electric charge density.
$\pd_t$ denotes the differentiation with respect to the time $t$ and
$\nabla$ is the Nabla operator.
In addition, the electric current density vector 
and the electric charge density
fulfill the continuity equation 
\begin{align}
\label{CE}
&\nabla\cdot \BJ+\pd_t\rho=0\,.
\end{align}
The constitutive equations for the fields in a vacuum 
(Maxwell-Lorentz relations) read
\begin{align}
\label{CE2}
\BD=\epsilon_0\, \BE\,,\qquad 
\BH=\frac{1}{\mu_0}\, \BB\,,
\end{align}
where $\epsilon_0$ is the vacuum permittivity and 
$\mu_0$ is the vacuum permeability.
The speed of light in vacuum is given by
\begin{align}
c^2=\frac{1}{\epsilon_0\mu_0}\,.
\end{align}

From the Maxwell equations~(\ref{ME-inh}) and (\ref{ME-h}),
inhomogeneous wave equations for the electromagnetic field strengths
follow
\begin{align}
\label{E-w}
\square\,\BE=-\frac{1}{\epsilon_0}\Big(\nabla\rho+\frac{1}{c^2}\, \pd_t\BJ\Big)
\end{align}
and
\begin{align}
\label{B-w}
\square\,\BB=\mu_0\,\nabla\times\BJ\,,
\end{align}
where the d'Alembert operator is defined by
\begin{align}
\square:=\frac{1}{c^2}\,\pd_{tt}-\Delta\,\qquad\text{with}\quad
\Delta=\nabla\cdot\nabla\ .
\end{align}

For zero initial conditions,
using the retarded Green function of the wave equation and some mathematical
manipulations,
the causal solutions of the inhomogeneous wave equations~(\ref{E-w}) and 
(\ref{B-w}) are given by retarded electromagnetic field
strength vectors:
\begin{align}
\label{E-J}
\BE(\rr,t)&=
\frac{1}{4\pi \epsilon_0}
\int_\VV \bigg(
\frac{\rho(\rr',t-R/c)}{R^3}\,\RR
+\frac{\pd_t \rho(\rr',t-R/c)}{c R^2}\,\RR
-\frac{\pd_t \BJ(\rr',t-R/c)}{c^2 R}\bigg)
 \d \rr'\,,\\
\label{B-J}
\BB(\rr,t)&=
\frac{\mu_0}{4\pi}
\int_\VV \bigg(
\frac{\BJ(\rr',t-R/c)}{R^3}
+\frac{\pd_t \BJ(\rr',t-R/c)}{c R^2}\bigg)\times \RR\ 
 \d \rr'\,,
\end{align}
where $\BR=\rr-\rr'$, $R=|\rr-\rr'|$,
$\rr\in\R^3$, $t\in\R$ and $\VV$ denotes the whole 3-dimensional space.
Eq.~(\ref{E-J}) is the time-dependent generalized Coulomb-Faraday law
and Eq.~(\ref{B-J}) is the time-dependent generalized Biot-Savart law 
(see also \citep{HM}).
Eqs.~(\ref{E-J}) and (\ref{B-J}) express the electromagnetic fields in
terms of their retarded sources $\rho$, $\BJ$, $\pd_t\rho$ and $\pd_t \BJ$
with full generality. 
They were originally derived by~\citet{Jefimenko} 
(see also~\citep{Jefimenko00,Jefimenko04}).
They appear also in the book of~\citet{CD} and in the third edition
of \citet{Lorrain}.
An equivalent representation was given by~\citet{PP} 
(see also~\citep{Jefimenko04}).
Both equations are nowadays called the Jefimenko equations 
in standard books on electrodynamics (e.g.~\citep{Jackson,Griffiths,HM}). 
They are fundamental, elegant
and very useful equations.

\section{A spatio-temporally varying electric current loop}

We investigate a spatio-temporally varying electric current loop.
We consider a closed loop of arbitrary shape (planar or non-planar) 
that can move arbitrary. 
The current density vector of a time-variable or  spatio-temporally varying
electric current loop at the time-dependent position $\Bs(t)$ 
is given by a line integral of the form 
\begin{align}
\label{J}
\BJ(\rr,t)&=\oint_{L(t)} I(t)\,\delta(\rr-\Bs(t))\, \d \BL(\Bs(t))\,,
\end{align}
where $L(t)$ is the loop curve at time $t$ 
and $\d \BL$ denotes the line element along the loop.
$I(t)$ is the time-dependent magnitude of the current.
In addition, there is no charge density, $\rho(\rr,t)=0$, and the current
vector fulfills $\nabla\cdot \BJ=0$.
For this particular situation, Eq.~(\ref{E-J}) simplifies to 
\begin{align}
\label{E-J2}
\BE(\rr,t)&=
-\frac{\mu_0}{4\pi }
\int_\VV \frac{\pd_t \BJ(\rr',t-R/c)}{R}\,
 \d \rr'\,,
\end{align}
which is the Faraday term of the generalized Coulomb-Faraday law~(\ref{E-J}).
\citet{Jefimenko00,Jefimenko04} called this term~(\ref{E-J2}) 
the electrokinetic field which is the electric field induced by 
a time-variable electric current.
Thus, the current distribution is non-stationary. 
For example, the current distribution is non-stationary even when the velocity
is constant or zero due to the time-dependent current magnitude $I(t)$.
A spatiotemporal current field may be achieved without any motion, 
with sources turned on and off at different times in different locations.
It can be seen in Eq.~(\ref{E-J2}) that a time-variable electric current
creates an electric field (anti)parallel to the current $(\pd_t \BJ$)
and the electrokinetic field exists only as long as the current is changing in
time.
Therefore, the electrokinetic field is different 
from the electrostatic field (see also~\citep{Jefimenko00,Jefimenko04}).
The current density vector appearing in Eqs.~(\ref{E-J2}) and (\ref{B-J})
is to be evaluated at the retarded time, which is less than the present time $t$. 
The retarded current density vector is given by
\begin{align}
\label{J-ret}
\BJ(\rr',\tr)&=\oint_{L(\tr)} I(\tr)\, \delta(\rr'-\Bs(\tr))\, 
\d\BL(\Bs(\tr))\,,
\end{align}
where $\Bs(\tr)$ here denotes the position of the loop 
at the retarded time
\begin{align}
\label{tr0}
\tr=t-|\rr-\rr'|/c=t-R/c\,.
\end{align}
Thus, substituting Eq.~(\ref{J-ret}) in Eqs.~(\ref{E-J2}) and (\ref{B-J}),
we obtain
\begin{align}
\label{E-HF0}
\BE(\rr,t)&=
-\frac{\mu_0}{4\pi }
\int_\VV \pd_t \oint_{L(\tr)} \frac{I(\tr)}{R}\,
\delta(\rr'-\Bs(\tr))\, \d \BL(\Bs(\tr))\, \d \rr'
\,,\\
\label{B-HF0}
\BB(\rr,t)&=
-\frac{\mu_0}{4\pi }
\int_\VV \oint_{L(\tr)} \frac{I(\tr)}{R^3}\,
\delta(\rr'-\Bs(\tr))\,\BR\times \d \BL(\Bs(\tr))\, \d \rr'
\nonumber\\
&\quad
-\frac{\mu_0}{4\pi }
\int_\VV \pd_t \oint_{L(\tr)} \frac{I(\tr)}{c\,R^2}\,
\delta(\rr'-\Bs(\tr))\,\BR\times \d \BL(\Bs(\tr))\, \d \rr'\,.
\end{align}

Here remain the integrals of the delta functions, which are done by changing 
the variable of integration from $\rr'$ to $\Bz=\rr'-\Bs(\tr)$ 
and using the Jacobian $J$ of this transformation 
(see, e.g., \citep{Jones,Eyges})
\begin{align}
J={\text{det}}\, \bigg(\frac{\pd\Bz}{\pd\rr'}\bigg)
=1-\frac{\BV(\tr)\cdot(\rr-\rr')}{c|\rr-\rr'|}\,,
\end{align}
\begin{align}
\int 
F(\rr')\,\delta(\rr'-\Bs(\tr))\, \d \rr'&=
\int F(\rr')\,\delta(\Bz)\,\frac{1}{J}\, \d \Bz\nonumber\\
&=\frac{F(\rr')}{J}\bigg|_{\Bz=0}
=\frac{F(\rr')}{1-\frac{\BV(\tr)\cdot(\rr-\rr')}{c|\rr-\rr'|}}
\bigg|_{\rr'=\Bs(\tr)}\,,
\end{align}
where $\BV=\dot{\Bs}$ is the velocity of every point of the current loop.
From the argument of the $\delta$-function we get $\rr'=\Bs(\tr)$. 
Therefore, now $\RR(\tr)=\rr-\Bs(\tr)$ and 
$R(\tr)=|\rr-\Bs(\tr)|$ are time-dependent
and the retarded time is now given by
\begin{align}
\label{tr2}
\tr=t-|\rr-\Bs(\tr)|/c=t-R(\tr)/c  \,,
\end{align}
where $\Bs(\tr)$ is the retarded position of the time-dependent source point.
Unfortunately, the retarded time $\tr(\rr,t)$ 
is not given directly,
but must be determined by solving Eq.~(\ref{tr2}) what may be quite
tedious. Only in some simple cases $\tr$ is easy to find 
(e.g. uniform motion).
If the loop is moving with velocity less than the speed of light, 
the solution of Eq.~(\ref{tr2}) is unique.
The retarded time is a result of the finite speed of 
the propagation for electrodynamic waves.
In addition, we define
\begin{align}
\label{P}
P(t')=R(t')-\BV(t')\cdot\BR(t')/c\,, 
\end{align}
and after the integration in $\rr'$, Eqs.~(\ref{E-HF0})
and (\ref{B-HF0}) become
\begin{align}
\label{E-HF}
\BE(\rr,t)&=
-\frac{\mu_0}{4\pi }\,
\pd_t \Bigg[\oint_{L(t')} \frac{I(t')}{P(t')}\,\d \BL(\Bs(t'))\Bigg]_{t'=\tr}
\,,\\
\label{B-HF}
\BB(\rr,t)&=
-\frac{\mu_0}{4\pi }
\Bigg[\oint_{L(t')} \frac{I(t')}{P(t')R^2(t')}\,
\BR(t')\times \d \BL(\Bs(t'))\Bigg]_{t'=\tr}
\nonumber\\
&\quad
-\frac{\mu_0}{4\pi c }\,
\pd_t \Bigg[\oint_{L(t')} \frac{I(t')}{P(t')R(t')}\,
\BR(t')\times \d \BL(\Bs(t'))\Bigg]_{t'=\tr}\,.
\end{align}
Eqs.~(\ref{E-HF}) and (\ref{B-HF}) have a remarkable structure. 
They are in some sense similar to the so-called Heaviside-Feynman formulas for 
a non-uniformly moving point charge~(see, e.g, \citep{Jackson,HM}). 
Since we investigate an electric current loop,
Eqs.~(\ref{E-HF}) and (\ref{B-HF}) have the structure of time-dependent 
line integrals depending on the retarded time $\tr$. 
Therefore, Eqs.~(\ref{E-HF}) and (\ref{B-HF}) are retarded
line integrals.
The fields~(\ref{E-HF}) and (\ref{B-HF}) must be evaluated at 
some earlier times $t'=\tr$ (the retarded times) and for the corresponding
point $\Bs(t')$ of the current loop.
Also the line element $\d\BL$ depends at every point $\Bs(t')$ on the
retarded time.

Because of the explicit dependence of the retarded times on  $\Bs(t')$,
every point $\Bs(t')$ on the loop $L(t')$ depends on its own retarded time. 
For a current loop, the time dependence of the electromagnetic fields is 
based on a retardation due to the retarded times which are functions of
the variables $\rr$, $\Bs$ and $t$.
The electromagnetic fields~(\ref{E-HF}) and (\ref{B-HF}) 
at the point $\rr$ and at time $t$ receive contribution from
every point $\Bs(\tr)$ on the electric current loop 
sending the electromagnetic waves at the retarded time $\tr$.

The electric and magnetic fields created by a non-stationary  
loop at the point of observation are the result of the signals sent out
by all the individual points $\Bs$ on the loop $L$ and simultaneously received
at the point of observation at the instant time $t$.
But different points on the loop are at different distances from the point of
observation, and the times needed for the signals originating 
from different points of the loop to arrive at the point of observation are 
different.
Thus, the signals that are received at the point of observation simultaneously 
at the time $t$ are sent out from different points of the loop 
at different retarded times~(\ref{tr2}). 
For a moving current loop 
these times are different not only due to different points
on the loop, but also because the loop moves.

The evaluation of the electromagnetic 
fields~(\ref{E-HF}) and (\ref{B-HF}) is not a
trivial task. Eqs.~(\ref{E-HF}) and (\ref{B-HF}) are 
complicated line integrals depending on the retarded times
and time derivatives outside the line integrals.
Since the loop is non-stationary and can move, the time derivative acts on the 
integrand as well as on the line element $\d\BL$. 
The evaluation of the time derivatives outside the integral
can be carried out if we use a transport 
theorem for a time-dependent line integral.
Using the transport theorem for a line integral~(\ref{Rel-L}),
Eqs.~(\ref{E-HF}) and (\ref{B-HF}) read 
\begin{align}
\label{E-HF2}
\BE(\rr,t)&=
-\frac{\mu_0}{4\pi }\,
 \Bigg[\oint_{L(t')}
\Bigg(
\bigg(
\pd_t\, \frac{I(t')}{P(t')}
+\BV(t')\cdot\overrightarrow{\nabla}_{\!\Bs(t')}\,  \frac{I(t')}{P(t')}
\bigg)\d \BL(\Bs(t'))
\nonumber\\
&\qquad\qquad\quad
+\frac{I(t')}{P(t')}
\BV(t') \, \overleftarrow{\nabla}_{\!\Bs(t')}\cdot \d \BL(\Bs(t'))\Bigg)
\Bigg]_{t'=\tr}
\end{align}
and 
\begin{align}
\label{B-HF2}
\BB(\rr,t)&=
-\frac{\mu_0}{4\pi }
\Bigg[\oint_{L(t')} \frac{I(t')}{P(t')R^2(t')}\,
\BR(t')\times \d \BL(\Bs(t'))\Bigg]_{t'=\tr}
\nonumber\\
&\quad
-\frac{\mu_0}{4\pi c }
 \Bigg[\oint_{L(t')}
\Bigg(
\bigg(
\pd_t\, \frac{I(t')\, \BR(t')}{P(t')\,R(t')}
+\BV(t')\cdot\overrightarrow{\nabla}_{\!\Bs(t')}\,   \frac{I(t')\, \BR(t')}{P(t')R(t')}
\bigg)\times \d \BL(\Bs(t'))
\nonumber\\
&\qquad\qquad\quad
+\frac{I(t')\, \BR(t')}{P(t')\,R(t')}\,
\times \BV(t') \, \overleftarrow{\nabla}_{\!\Bs(t')}\cdot \d \BL(\Bs(t'))\Bigg)
\Bigg]_{t'=\tr}\,.
\end{align}
In Eqs.~(\ref{E-HF2}) and (\ref{B-HF2}), the derivative with respect to the
present time $t$ and 
derivatives with respect to the time-dependent source point of the loop 
$\Bs(t')$, acting on the corresponding retarded time $t'=\tr$, 
appear. The derivatives with respect to the time-dependent source points of the
loop are a consequence that we investigate an electric
current loop instead of an electric  point charge.

To obtain the electromagnetic fields in a `Li\'enard-Wiechert' type form,
we must carry out the time derivatives and the derivatives with respect 
to the time-dependent source point  $\Bs(t')$, 
which are not trivial because of the subtle relation between present and retarded times.
Using the corresponding formulas for the 
derivatives~(\ref{dt-I})--(\ref{dt-PR}) and ~(\ref{ds-I})--(\ref{ds-PR}),
we find
\begin{align}
\label{E-LW}
\BE(\rr,t)&=
-\frac{\mu_0}{4\pi }\,
\Bigg[
\oint_{L(t')}
\Bigg(
\dot{I}(t')\bigg(\frac{R(t')}{P^2(t')}+ \frac{\BV(t')\cdot\BR(t')}{c\, P(t')\,R(t')}\bigg)
\nonumber\\
&\qquad
+\frac{I(t')}{P^3(t')}
\bigg(\Big(\dot{\BV}(t')\cdot\BR(t') -V^2(t')\Big)\frac{R(t')}{c}
+\BV(t')\cdot\BR(t')\bigg)
\nonumber\\
&\qquad
+\frac{I(t')}{P^2(t')R(t')}
\bigg(\dot{\BV}(t')\cdot\BR(t')\, \frac{{\BV}(t')\cdot\BR(t')}{c^2}
 -\frac{R(t')\, V^2(t')}{c} +\BV(t')\cdot\BR(t')\bigg)
\Bigg)\d \BL(\Bs(t'))
\nonumber\\
&\qquad
+\oint_{L(t')}\frac{I(t')}{P(t')}\,
\frac{\dot{\BV}(t')}{c\, R(t')}\, \BR(t')\cdot \d \BL(\Bs(t'))
\Bigg]_{t'=\tr}
\end{align}
and
\begin{align}
\label{B-LW}
\BB(\rr,t)&=
-\frac{\mu_0}{4\pi c}\,
\Bigg[
\oint_{L(t')}
\Bigg(
\dot{I}(t')\bigg(\frac{1}{P^2(t')}+ \frac{\BV(t')\cdot\BR(t')}{c\, P(t')\,R^2(t')}\bigg)
\nonumber\\
&\qquad
+\frac{I(t')}{P^3(t')R(t')}
\bigg(\Big(\dot{\BV}(t')\cdot\BR(t') -V^2(t')\Big)\frac{R(t')}{c}
+\BV(t')\cdot\BR(t')\bigg)
\nonumber\\
&\qquad
+\frac{I(t')}{P^2(t')R^2(t')}
\bigg(\dot{\BV}(t')\cdot\BR(t')\, \frac{{\BV}(t')\cdot\BR(t')}{c^2}
 -\frac{R(t')\, V^2(t')}{c} +\BV(t')\cdot\BR(t')\bigg)
\nonumber\\
&\qquad
+I(t')\,\BV(t')\cdot\BR(t')\bigg(\frac{1}{P^2(t')R^2(t')}+\frac{1}{P(t')R^3(t')}\bigg)\nonumber\\
&\qquad
+\frac{c\, I(t')}{P(t')R^2(t')}\Bigg)\BR(t')\times\d \BL(\Bs(t'))
\nonumber\\
&\qquad
-\oint_{L(t')} I(t')\, \bigg(\frac{1}{P^2(t')}+\frac{1}{P(t')R(t')}\bigg)
\BV(t')\times\d \BL(\Bs(t'))
\nonumber\\
&\qquad
+\oint_{L(t')} \frac{I(t')}{P(t')}\,
\frac{\BR(t')\times\dot{\BV}(t')}{c\, R^2(t')}\, \BR(t')\cdot \d \BL(\Bs(t'))
\Bigg]_{t'=\tr}\,.
\end{align}
These are the electromagnetic fields induced by the
spatio-temporally varying electric current loop. 
These expressions completely describe the electromagnetic fields of the
spatio-temporally varying electric current loop.
The expressions for the electromagnetic fields of the 
electric current loop are given in terms of 
the retarded positions, retarded velocities, and the retarded line element. 
For a general non-uniform motion the relation of the retarded position 
to the present position is not known.
Eq.~(\ref{E-LW}) is the generalized Faraday law or electrokinetic field of 
a spatio-temporally varying electric current loop 
and Eq.~(\ref{B-LW}) is the generalized Biot-Savart law for
a spatio-temporally varying electric current loop.
There are two sources for the induction of the electromagnetic fields:
the time-dependence of the current magnitude and 
the fact that positions of the electric current loop are time-dependent.

If the current loop is moving uniformly, which means with constant
velocity $\BV$ and zero acceleration, the induced electromagnetic fields 
are obtained from Eqs.~(\ref{E-LW}) and (\ref{B-LW})
putting $\dot{\BV}=0$.
The solution of the uniform motion of a current loop is more complicated 
than the usual steady-state solution of a point charge 
(see, e.g.,~\citep{PP,Griffiths}), 
due to the uniformly moving line element of the current loop. 
One non-trivial task for the uniform motion is to express the electromagnetic
fields in terms of the so-called present position: 
$\RR_0=\big[\RR(\tr)-R(\tr) \BV/c\big]$ where $\tr$ denotes the 
retarded time for the uniform motion\footnote{
For the uniform motion, the retarded time reads~(see, e.g.,~\citep{Griffiths})
\begin{align*}
\tr=\frac{(c^2 t-\BV\cdot \rr)-\sqrt{\big(c^2 t-\BV\cdot \rr\big)^2
+\big(c^2-V^2\big)\big(\rr^2-c^2t^2\big)}}{c^2-V^2}\,,
\end{align*}
which is the solution of Eq.~(\ref{tr2}) for $\Bs(t)=\BV t$ and 
$\BV=\text{constant}$. 
}.

Since the time derivative of the line integrals in Eqs.~(\ref{E-HF}) and 
(\ref{B-HF}) is involved, it is clear that the electromagnetic
fields~(\ref{E-HF}) and (\ref{B-HF}) 
will be functions not only of the velocity~$\BV$,
but also of the acceleration~$\dot{\BV}$.  
We may therefore separate $\BE$ and $\BB$ into two parts each, one 
which involves the acceleration and  
the time-derivative of the magnitude of the current
and goes to zero for $\dot{\BV}=0$ and $\dot I=0$,
and one which involves only the velocity $\BV$:
\begin{align}
\label{E-dec}
\BE(\rr,t)&=\BE_\text{nonrad}(\rr,t)+\BE_\text{rad}(\rr,t)\,,\\
\BB(\rr,t)&=\BB_\text{nonrad}(\rr,t)+\BB_\text{rad}(\rr,t)\,.
\end{align}
The fields 
$\BE_\text{nonrad}$ and $\BB_\text{nonrad}$ are called 
nonradiation parts and the fields $\BE_\text{rad}$ and $\BB_\text{rad}$ 
are called the radiation parts.
The electric parts are
\begin{align}
\label{E-nonr}
\BE_\text{nonrad}(\rr,t)&=
\frac{\mu_0}{4\pi }\,
\Bigg[
\oint_{L(t')}
\Bigg(\frac{I(t')}{P^3(t')}
\bigg(\frac{V^2(t')\, R(t')}{c}-\BV(t')\cdot\BR(t')\bigg)
\nonumber\\
&\qquad\qquad
+\frac{I(t')}{P^2(t')R(t')}
\bigg(\frac{V^2(t')\, R(t')}{c}-\BV(t')\cdot\BR(t')\bigg)
\Bigg)\d \BL(\Bs(t'))
\Bigg]_{t'=\tr}
\end{align}
\begin{align}
\label{E-r}
\BE_\text{rad}(\rr,t)&=
-\frac{\mu_0}{4\pi }\,
\Bigg[\oint_{L(t')}
\Bigg(
\dot{I}(t')\bigg(\frac{R(t')}{P^2(t')}
+\frac{\BV(t')\cdot\BR(t')}{c\, P(t')\,R(t')}\bigg)
\nonumber\\
&\qquad\qquad
+I(t')\,\dot{\BV}(t')\cdot\BR(t')
\bigg(\frac{R(t')}{c\, P^3(t')}+\frac{\BV(t')\cdot\BR(t')}{c^2\, P^2(t')\, R(t')}
\bigg)\Bigg)\d \BL(\Bs(t'))
\nonumber\\
&\qquad\quad
+\oint_{L(t')}\frac{I(t')\,\dot{\BV}(t')}{c\, P(t')\,R(t')}\, 
\BR(t')\cdot \d \BL(\Bs(t'))
\Bigg]_{t'=\tr}
\end{align}
and the magnetic parts read
\begin{align}
\label{B-nonr}
\BB_\text{nonrad}(\rr,t)&=
\frac{\mu_0}{4\pi c}\,
\Bigg[
\oint_{L(t')}
\Bigg(\frac{I(t')}{P^3(t')R(t')}
\bigg(\frac{V^2(t')\, R(t')}{c}-\BV(t')\cdot\BR(t')\bigg)
\nonumber\\
&\qquad\qquad
+\frac{I(t')}{P^2(t')R^2(t')}
\bigg(\frac{V^2(t')\, R(t')}{c}-\BV(t')\cdot\BR(t')\bigg)
\nonumber\\
&\qquad\qquad
-I(t')\,\BV(t')\cdot\BR(t')\bigg(\frac{1}{P^2(t')R^2(t')}+\frac{1}{P(t')R^3(t')}\bigg)\nonumber\\
&\qquad\qquad
-\frac{c\,I(t')}{P(t')R^2(t')}\Bigg)\BR(t')\times\d \BL(\Bs(t'))
\nonumber\\
&\qquad
+\oint_{L(t')} I(t')\, \bigg(\frac{1}{P^2(t')}+\frac{1}{P(t')R(t')}\bigg)
\BV(t')\times\d \BL(\Bs(t'))
\Bigg]_{t'=\tr}
\end{align}
\begin{align}
\label{B-r}
\BB_\text{rad}(\rr,t)&=
-\frac{\mu_0}{4\pi c}\,
\Bigg[\oint_{L(t')}
\Bigg(
\dot{I}(t')\bigg(\frac{1}{P^2(t')}
+\frac{\BV(t')\cdot\BR(t')}{c\, P(t')\,R^2(t')}\bigg)
\nonumber\\
&\qquad\qquad
+I(t')\,\dot{\BV}(t')\cdot\BR(t')
\bigg(\frac{1}{c\, P^3(t')}+\frac{\BV(t')\cdot\BR(t')}{c^2\, P^2(t')\, R^2(t')}
\bigg)\Bigg)\BR(t')\times\d \BL(\Bs(t'))
\nonumber\\
&\qquad\quad
+\oint_{L(t')} I(t')\,\frac{\BR(t')\times\dot{\BV}(t')}{c\, P(t')\,R^2(t')}\, 
\BR(t')\cdot \d \BL(\Bs(t'))
\Bigg]_{t'=\tr}\,.
\end{align}
The nonradiation parts~(\ref{E-nonr}) and (\ref{B-nonr}) are
formally the `convective' fields of a uniformly moving current loop at the 
retarded position. 
In addition, it can be seen in Eqs.~(\ref{E-r}) and (\ref{B-r}) that
the electromagnetic radiation fields produced by 
the spatio-temporally varying
electric current loop are more  
complicated than the radiation fields of a point 
charge (see, e.g., \citep{PP,Jackson,Griffiths,HM}).
If the loop is accelerated, the electromagnetic fields are neither
static nor convective, and there is a net change in the field energy which
causes radiation. 
The electromagnetic radiation possesses two sources and 
is caused by the time-change of the 
magnitude of the current loop $\dot{I}$ and the acceleration $\dot\BV$.

From these equations, it is apparent that the static limit 
(i.e., $\BV=0$, $\dot{\BV}=0$, $I=\text{constant}$) is satisfied, since
\begin{align}
\BE(\text{static})
&=\BE_\text{nonrad}(\text{static})+\BE_\text{rad}(\text{static})\nonumber\\
&=0+0\\
\BB(\text{static})
&=\BB_\text{nonrad}(\text{static})+\BB_\text{rad}(\text{static})\nonumber\\
&=\BB_\text{nonrad}(\text{static})+0\nonumber\\
\label{B-stat}
&=-\frac{\mu_0 I}{4\pi}\oint_L\frac{1}{R^3}\, \BR\times\d \BL'\,,
\end{align}
which is the appropriate expression for the Biot-Savart law 
(see, e.g.,~\citep{Griffiths}).

We note the following radial dependencies of the electromagnetic fields:
\begin{align}
\BE_\text{nonrad},\ \BB_\text{nonrad}\  
&\simeq \bigg[\oint_L \frac{1}{R^2}\,\d \BL
\bigg]_{\tr}
\\
\BE_\text{rad},\ \BB_\text{rad}\  &\simeq \bigg[\oint_L \frac{1}{R}\,\d \BL
\bigg]_{\tr}\,.
\end{align}
The nonradiation fields $\BE_\text{nonrad}$ and 
$\BB_\text{nonrad}$ are the near fields since they
fall off as $1/R^2$ in the integrand. 
The radiation or acceleration fields $\BE_\text{rad}$ and $\BB_\text{rad}$ 
are the far fields falling off as 
$1/R$ in the integrand. The radiation fields then dominate at large distances
and they must be used to calculate energy loss by radiation.
As an application, the obtained results can be used for the 
radiation of non-uniformly moving loop antennas.

Another important limit of our general results, is the 
case of a current loop with time-variable magnitude $I(t)$ 
and localized at the position $\rr'$ with fixed loop shape $L$.
The corresponding current density vector reads
\begin{align}
\label{J-2}
\BJ(\rr,t)&=\oint_{L} I(t)\,\delta(\rr-\rr')\, \d \BL'\,.
\end{align}
Since the position of the loop is fixed at the point $\rr'$,
the velocity and the acceleration of the positions of the loop are 
zero $\BV=0$ and $\dot{\BV}=0$.
The electric and magnetic fields can be obtained directly 
from Eqs.~(\ref{E-LW}) and (\ref{B-LW}), or by substituting 
Eq.~(\ref{J-2}) into the Jefimenko equations~(\ref{E-J2}) and (\ref{B-J}).
The electric field (electrokinetic field) is given by
\begin{align}
\label{E-L}
\BE(\rr,t)&=
-\frac{\mu_0}{4\pi }\,
\oint_{L}
\frac{\dot{I}(t-R/c)}{R}\, \d\BL'\,,
\end{align}
and the magnetic field is
\begin{align}
\label{B-L}
\BB(\rr,t)&=
-\frac{\mu_0}{4\pi }\,
\oint_{L}
\bigg(
\frac{I(t-R/c)}{R^3}+
\frac{\dot{I}(t-R/c)}{c\,R^2}\bigg)\BR\times \d\BL'\,,
\end{align}
which are in agreement with the result given earlier by~\citet{Jefimenko}.

\section*{Acknowledgement}
The author gratefully acknowledges the grants from the 
Deutsche Forschungsgemeinschaft (Grant Nos. La1974/2-1, La1974/3-1). 
The author wishes to express his gratitude to 
Prof. Wolfgang Ellermeier for stimulating discussions and useful remarks.

\begin{appendix}
\setcounter{equation}{0}
\renewcommand{\theequation}{\thesection.\arabic{equation}}
\section{Appendix 
}
For the convenience of the reader, we give all the formulas in the appendix in
the index notation.
\label{appendixA}
\subsection{Generalized transport theorem for a line integral}
The generalized transport theorem for a line integral 
reads~(see, e.g., \citep{Eringen,Chadwick})
\begin{align}
\label{Rel-L}
\pd_t \oint_{L(t)} f(\rr,\rr',t)\, \d L'_i =
 \oint_{L(t)}\bigg[\Big(\pd_t f+ V'_{j} \pd_{j'}f\Big)\d L'_i
+f\,  \pd_{j'}V'_{i}\, \d L'_j
\bigg]\,,
\end{align}
where $\BV'=\dot{\rr}'$ is the velocity of every point of the moving line
(drift velocity). The derivative $\pd_t +V'_{j} \pd_{j'}$ 
on the right hand side is often called the co-moving time derivative.
$\d\BL'$ is a time-dependent line element.
The last term in Eq.~(\ref{Rel-L}) is due to the time-derivative of the 
line element and contains the velocity gradient.
In general, $f$ may be a tensor-valued function.

\subsection{Derivatives at the retarded time}
Here we give some useful relations of derivatives of quantities 
depending on the retarded time, which is the unique solution of the 
relation
\begin{align}
t-\tr-|\rr-\Bs(\tr)|/c=0\,.
\end{align}
First, we carry out the time derivatives, which are not
trivial because of the subtle relation between present and retarded time
(see also~\citep{Barton}):
\begin{align}
\label{dt-t}
 \bigg[\frac{\pd t'}{\pd t}\bigg]_{t'=\tr}
=\bigg[\frac{R(t')}{P(t')}\bigg]_{t'=\tr}
\end{align}
\begin{align}
\label{dt-I}
\pd_t \big[I(t')\big]_{t'=\tr}
=\bigg[\frac{\pd t'}{\pd t}\, \frac{\pd I(t')}{\pd t'}\bigg]_{t'=\tr}
=\bigg[\frac{R(t')}{P(t')}\, \dot{I}(t')\bigg]_{t'=\tr}
\end{align}
\begin{align}
\label{dt-P}
\pd_t \bigg[\frac{1}{P(t')}\bigg]_{t'=\tr}
=\bigg[\frac{1}{P^3(t')}\bigg(
\big(\dot{V}_m(t') R_m(t') -V^2(t')\big)\, 
\frac{R(t')}{c}+V_m(t') R_m(t')\bigg)\bigg]_{t'=\tr}
\end{align}
\begin{align}
\label{dt-PR}
\pd_t \bigg[\frac{R_k(t')}{P(t')R(t')}\bigg]_{t'=\tr}
&=\bigg[\frac{R_k(t')}{P^3(t')R(t')}\bigg(
\Big(\dot{V}_m(t') R_m(t') -V^2(t')\Big)\, \frac{R(t')}{c}
+V_m(t') R_m(t')\bigg)
\nonumber\\
&\qquad 
+\frac{1}{P^2(t')}\bigg(
V_m(t') R_m(t')\, \frac{R_k(t')}{R^2(t')}-V_k(t')\bigg)\bigg]_{t'=\tr}\,.
\end{align}

Secondly, we give the relations for the derivatives
of the quantities depending on the retarded time
with respect to the retarded position of 
moving source point (see also~\citep{Jefimenko04}):
\begin{align}
\label{ds-R}
 \bigg[\frac{\pd R_i(t')}{\pd s_j(t')}\bigg]_{t'=\tr}
=-\delta_{ij}
\end{align}
\begin{align}
\label{ds-t}
 \bigg[\frac{\pd t'}{\pd s_j(t')}\bigg]_{t'=\tr}
=\bigg[\frac{R_j(t')}{c\, R(t')}\bigg]_{t'=\tr}
\end{align}
\begin{align}
\label{ds-I}
 \bigg[\frac{\pd}{\pd s_j(t')}\,  I(t')\bigg]_{t'=\tr}
=\bigg[\frac{\pd t'}{\pd s_j(t') }\, \frac{\pd I(t')}{\pd t'}\bigg]_{t'=\tr}
=\bigg[\frac{R_j(t')}{c\, R(t')}\, \dot{I}(t')\bigg]_{t'=\tr}
\end{align}
\begin{align}
\label{ds-V}
 \bigg[\frac{\pd}{\pd s_j(t')}\,  V_i(t')\bigg]_{t'=\tr}
=\bigg[\frac{R_j(t')}{c\, R(t')}\, \dot{V}_i(t')\bigg]_{t'=\tr}
\end{align}
\begin{align}
\label{ds-P}
 \bigg[\frac{\pd}{\pd s_j(t')}\, \frac{1}{P(t')}\bigg]_{t'=\tr}
&=\bigg[\frac{1}{P^2(t') R(t')}\bigg(R_j(t')
-\frac{R(t')V_j(t')}{c}
+\dot{V}_m(t') R_m(t')\, \frac{R_j(t')}{c^2}
\bigg)
\bigg]_{t'=\tr}
\end{align}
\begin{align}
\label{ds-PR}
 \bigg[\frac{\pd}{\pd s_j(t')}\, 
\frac{R_k(t')}{P(t')R(t')}\bigg]_{t'=\tr}
&=
\bigg[\frac{R_k(t')}{P^2(t') R^2(t')}
\bigg(R_j(t')-\frac{R(t')V_j(t')}{c}
+\dot{V}_m(t') R_m(t')\, \frac{R_j(t')}{c^2}\bigg)\nonumber\\
&\qquad
-\frac{\delta_{jk}}{P(t') R(t')}
+\frac{R_j(t')R_k(t')}{P(t') R^3(t')}\bigg]_{t'=\tr}\,.
\end{align}

\end{appendix}

\end{document}